\documentclass[3p]{elsarticle} 
\usepackage{amsmath} 
\usepackage{amssymb} 
\usepackage{subfigure}

\usepackage{amsfonts} \journal{}

\usepackage{graphicx,color} 
\usepackage{multirow,bigstrut}

\graphicspath{{imgs/}}

\usepackage{psfrag}

\usepackage{amssymb,amsmath,bm}
\newcommand\mymatrix[1]{\bm{\mathrm{#1}}} 
\usepackage{url}

\definecolor{dgreen}{rgb}{0,.6,0}

\usepackage[normalem]{ulem} 
\usepackage{hyperref} 
\begin{document}

\newlength\figwidth \setlength\figwidth{0.5\columnwidth}
\newlength\imgwidth \setlength\imgwidth{0.3\columnwidth}

\begin{frontmatter} \title{Cryptanalysis of a one round chaos-based Substitution Permutation Network}
\author[Spain]{David Arroyo\corref{corr}}
\author[Spain]{Jesus Diaz}
\author[Spain]{F.B. Rodriguez} 
\cortext[corr]{Corresponding author: David Arroyo (david.arroyo@uam.es).}  
\address[Spain]{Grupo de Neurocomputaci\'on Biol\'ogica, Dpto. de Ingenier\'ia
Inform\'atica. Escuela Polit\'ecnica Superior. Universidad Aut\'onoma
de Madrid, 28049 Madrid, Spain }

\begin{abstract} The interleaving of chaos and cryptography has been
the aim of a large set of works since the beginning of the
nineties. Many encryption proposals have been introduced to
improve conventional cryptography. However, many of those
proposals possess serious problems according to the basic requirements
for the secure exchange of information. In this paper we highlight
some of the main problems of chaotic cryptography by means of the
analysis of a very recent chaotic cryptosystem based on a
one round Substitution Permutation Network. More specifically, we
show that it is not possible to avoid the security problems
of that encryption architecture just by including a chaotic system as
core of the derived encryption system.
\end{abstract}
\end{frontmatter}

\section{Introduction}

\label{sec:intro} The plinth of cryptography is built upon the
properties of confusion and diffusion as stated by Shannon in 1949
\cite{shannon49}, which can be linked to the main characteristics of
chaotic systems: ergodicity and sensitivity to control parameters and
initial conditions. The connection between the basic coordinates of
cryptography and chaotic systems has paved the research on chaotic
cryptography \cite{book}. A lot of different methods have been proposed
in the field of chaos-based cryptography, but most of them show very
serious security flaws \cite[Chapters 8 and 9]{book}. A very important
family of chaotic cryptosystems is the one inheriting the
characteristics of the Substitution-Permutation Networks (SPNs), as
it is explained in \cite{Fridrich1998}. This kind of architecture is
not secure unless the avalanche criterion is satisfied
\cite{Tavares1995}. As matter of fact, the inclusion of chaotic
systems in this kind of architecture does not guarantee security and
the assessment of the avalanche property should be thoroughly carried
out \cite{Solak2010,arroyo08b}. In \cite{Wang20121101} a chaotic
cryptosystem is proposed to encrypt colour images through the
permutation of their columns and rows, along with a substitution
procedure based on the logistic map. From a general point of view,
this cryptosystem can be interpreted as one round of a SPN. This kind
of architecture present a very low level of confusion and security
pitfalls if the substitution stage can be rewritten as a way to change
the plaintext according to a keystream which is independent of the
plaintext. As we will discuss along this paper, this is the case of
the cryptosystem described in \cite{Wang20121101}.

The rest of the paper is organized as follows. In
Sec.~\ref{sec:description} it is described the cryptosystem under
examination. With the aim of underlining the shortcomings of this
encryption scheme, we discuss in Sec.~\ref{sec:weaknesses}
some limitations with respect to the 
dynamical system bearing encryption, to the key space and, finally, in
regards to the diffusion  property of the cryptosystem. The analysis
is complemented by remarking the vulnerability of the cryptosystem
against a chosen-plaintext attack. In this concern, we explain along
Sec.~\ref{sec:probl-deriv-from} how to elude the security laying on
the encryption architecture selected in \cite{Wang20121101}. Finally,
in Sec.~\ref{sec:conclusions} we summarize and discuss the results of
the cryptanalysis. 

 \begin{figure}[!htbp]
  \centering
  \includegraphics[scale=0.8]{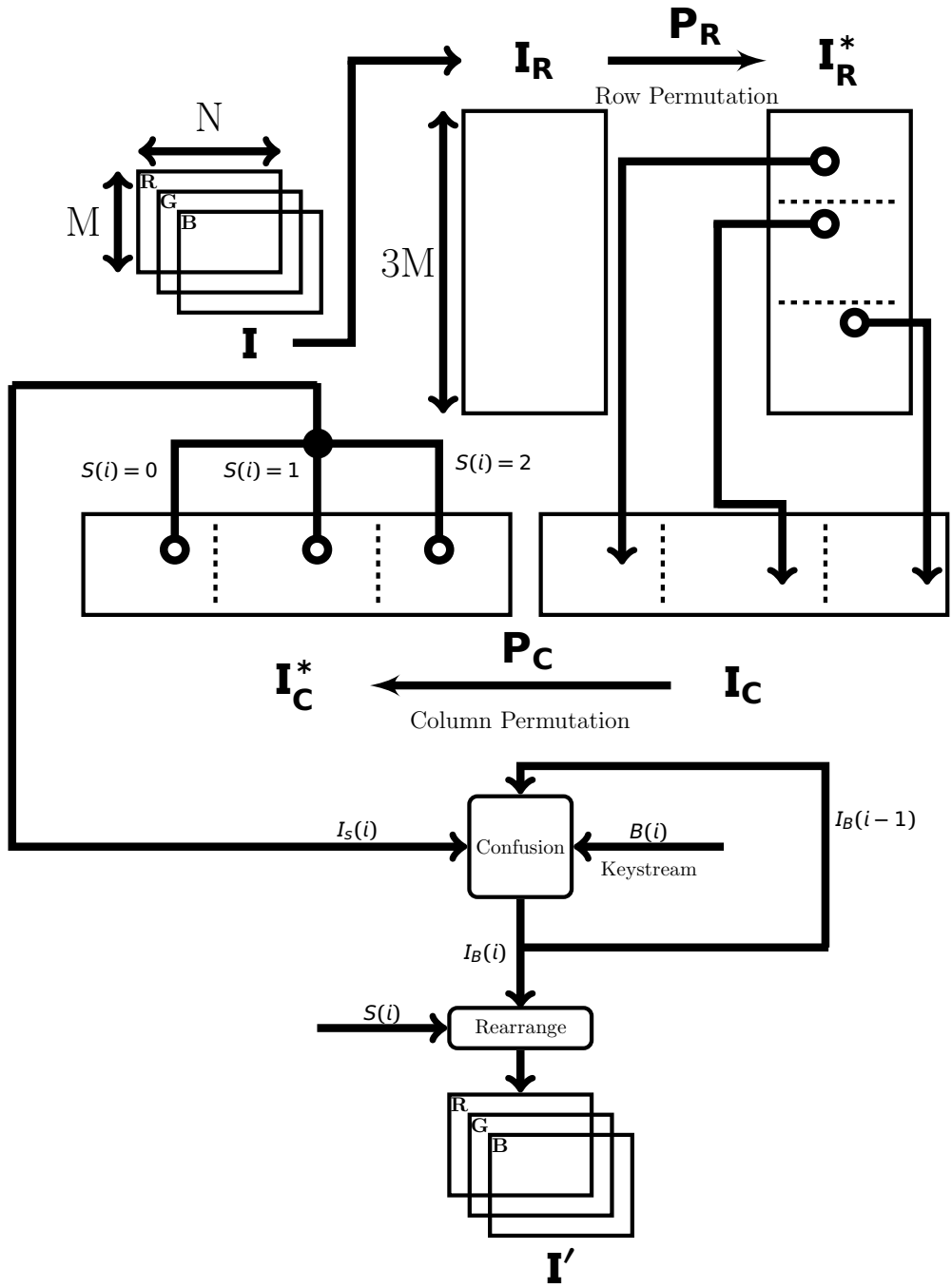}
  \caption{Diagram of the encryption procedure.}
  \label{fig:encryption}
\end{figure}
\section{Description of the encryption scheme}
\label{sec:description} The encryption procedure defined in
\cite{Wang20121101} is applied on colour plain-images of size $M
\times N$ and coded in RGB format. The plain colour image is treated
as a matrix  $\mymatrix{I}$ of size $M\times N \times 3$, whereas the
cipher-image is given by $\mymatrix{I}'$ also of size $M\times N \times 3$. For the sake of clarity, we have first
modified the notation used in \cite{Wang20121101} and second divided
the encryption method into four stages (see Fig.~\ref{fig:encryption}):

\begin{enumerate}
  \item \emph{Rows permutations}.

    The colour plain-image $\mymatrix{I}$ is transformed into a
gray-scale image $\mymatrix{I_R}$ of size $3M\times N$, just by
incorporating the rows of the green and blue components after the rows
of the red one. Let $\mymatrix{P}_R$ be a permutation matrix that
transforms $\mymatrix{I_R}$ into $\mymatrix{I}_R^*$ by shuffling its
rows in a pseudo-random way, through the iteration of the logistic map
for control parameter equals to $\lambda_R$ and initial condition
given by $x_R$. The logistic map is defined by the iteration function
    \begin{equation}
      \label{eq:logistic} f(x) = \lambda x (1-x),
    \end{equation} and the orbit $\left\{x(i)\right\}_{i}$ can be
generated from a given initial condition $x(0)$ by doing
$x(i+1)=f(x(i))$.

  \item \emph{Columns permutations}.
    
    The matrix $\mymatrix{I}_R^*$ is converted into a matrix
$\mymatrix{I_C}$ of size $M\times 3N$, by combining horizontally (one
after the other) the three groups of $M$ rows that define
$\mymatrix{I}_R^*$. For each row of $\mymatrix{I_C}$, the pixels are
permuted according to the corresponding row of a permutation matrix
$\mymatrix{P_C}$. The resulting matrix is noted as
$\mymatrix{I}_C^{*}$. Again, this permutation matrix is obtained by
iterating the logistic map in this case with control parameter $\lambda_C$ and
initial condition $x_C$.

  \item \emph{Selection of the next pixel to encrypt}.
    
    Once pixels have been shuffled, substitution is performed using a
keystream and selecting the pixel to encrypt based on a pseudo-random
sequence $\left\{S(i)\right\}_{i=1}^{3MN}$, with $S(i) \in
\left\{0,1,2\right\}$. The sequence
$\left\{S\right\}_{i=1}^{3MN}$ determines if the next pixel to encrypt
proceeds from either the first N columns ($S(i)=0$), the second group of N columns ($S(i)=1$), or the
third set of columns (for $S(i)=2$) of $\mymatrix{I_C}^*$. In case all pixels of a
band have been already selected, the pixel to encrypt is chosen from
the next colour band (after the blue pixels, the next ones are the
red). Consequently, a vector $\mymatrix{I}_S$ of length $3MN$ is
obtained by reading each colour component of $\mymatrix{I}_C^*$ from the
first row and from the left to the right, according to the selection
vector $\mymatrix{S}$.

  \item \emph{Substitution stage}.

    Finally, the output of the previous step is masked using a
keystream $\left\{B(i)\right\}_{i=1}^{3MN}$. The update rule is given
by
    \begin{equation} I_B(i) = \left\{\begin{array}{lr}(I_S(i)+
B(i))mod \ 256,& i=1 \\ (I_B(i-1) + I_S(i)+ I_S(i-1) + B(i)) mod \
256,& i=2\sim3MN\end{array}\right.
\label{eq:1}
    \end{equation} 
The resulting cipher-image $\mymatrix{I}'$ is
derived from $\mymatrix{I_B}$ using $\mymatrix{S}$, i.e., by grouping
the pixels of $\mymatrix{I}_B$ into colour components in the reversed
order that they were grabbed from $\mymatrix{I}_C^*$ to build up
$\mymatrix{I}_S$.
\end{enumerate}

According to \cite{Wang20121101}, the secret key of the cryptosystem
consists of the set of values
$\left\{\lambda_R,x_R,\lambda_C,x_C\right\}$, which are used to compute
two orbits of the logistic map (Eq.~\eqref{eq:logistic}). Those orbits
are the core of the procedures to generate the permutation matrices
$\mymatrix{P}_R$ and $\mymatrix{P}_C$, the pseudo-random sequence
$\mymatrix{S}$, and the keystream $\mymatrix{B}$. As we discuss below,
the cryptanalysis of the cryptosystem can be carried out independently
of those generation procedures. For a more detailed description of any
of those procedures or other design details, please refer to
Sec. 2.1 of \cite{Wang20121101}.

\section{Design weaknesses}
\label{sec:weaknesses} As result of our previous work on the field of
chaos-based cryptography \cite{arroyo:thesis}, we can conclude that
the most critical problems in chaotic cryptography are linked to three
aspects: the selection of the chaotic system, the choice of an
encryption architecture, and the implementation of the
cryptosystem. In the specific scenario depicted by
\cite{Wang20121101}, there exist some problems that we have previously
highlighted in regards to both the selection of the chaotic system and
the encryption architecture
\cite{arroyo08a,arroyo10,arroyo_recsi_2008,arroyo_recsi_2010,arroyo08b}. Those
problems inform about a non exhaustive description of the
cryptosystem, but also about security breaches. The drawbacks of the
cryptosystem definition are derived in Sec.~\ref{sec:non-exha-defin} by studying the key space of the
cryptosystem on account of the dynamical properties
of the underlying chaotic map, and in
Sec.~\ref{sec:low-sens-change} through the discussion of the diffusion property of the encryption
architecture. The security analysis is the core of Sec.~\ref{sec:probl-deriv-from}.

\subsection{Non exhaustive definition of the key space}
\label{sec:non-exha-defin} 
\begin{figure}[!htbp]
  \centering
  \includegraphics[scale=0.7]{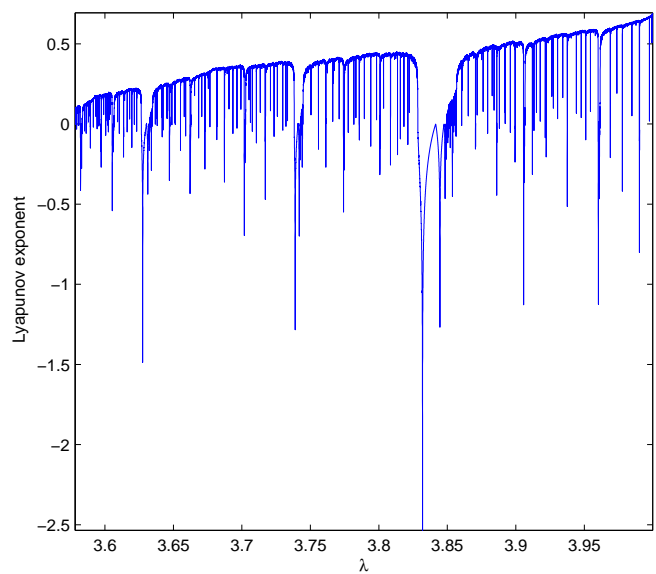}
  \caption{Lyapunov exponent of the logistic map with respect to the
    control parameter $\lambda$. The selection of $\lambda_R$ and $\lambda_C$ should be performed guaranteeing chaoticity, i.e.,
    positive values for the Lyapunov exponent.}
  \label{fig:lyapunov}
\end{figure}

One major concern in chaotic cryptography
is on designing cryptosystems in such a way that the underlying
dynamical systems evolves chaotically \cite[Rule 5]{Alvarez06a}. In the
case of the logistic map (and other maps), this resorts to the
evaluation of the Lyapunov exponent in order to guarantee
chaoticity (see Fig.~\ref{fig:lyapunov}). As a matter of fact, after the Myrberg-Feigenbaum point
($\lambda\approx 3.5699456$) it cannot be asserted that the logistic
map is always chaotic due to the existence of a dense set of periodic
windows (i.e., of values of $\lambda$ implying \emph{regular} and
\emph{non stochastic} behavior \cite{Tucker20091923}).

\begin{figure}[!htbp] \centering
  \includegraphics[scale=0.7]{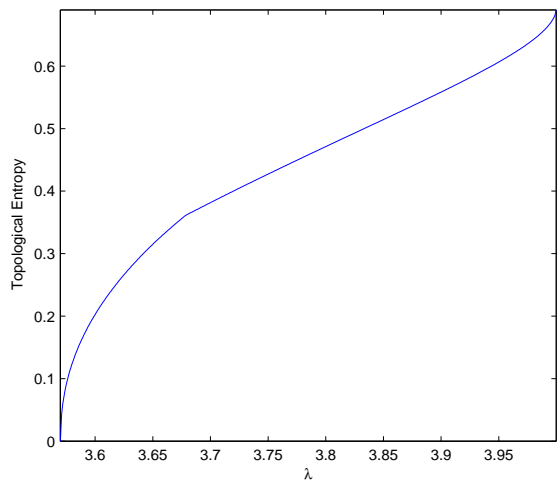}
  \caption{Topological entropy of the logistic map.}
  \label{fig:top_log}
\end{figure}

Additionally, in \cite{Wang20121101} the use of the logistic map relies
not only on its positive rate of local divergence, but also on its
topological properties. Certainly, the permutation of columns and rows
is conducted by the ordering of chaotic orbits of the logistic map of
length $3MN$ and $3M$, respectively. In this sense, we should assess
whether the number of possible permutations on the values of those
orbits is at least equal to the number of possible initial
conditions. The number of initial conditions is given by the inverse
of the \emph{machine epsilon} \cite[p. 37]{floatingPoint:higham},
which is $2^{52}$ for double precision floating-point arithmetic.  On
the other hand,  the number of possible permutations on a given orbit
of length $L$  is $L!$. In the case of deterministic dynamical
systems, this upper value is not reached due to the existence of a set of
\emph{forbidden} permutations \cite{amigo07}. If we restrict our
discussion to dynamical systems with iteration function
$f_{\lambda}$ defined as a
scalar, then the cardinality of the set of possible permutations of an orbit is upper bounded
by $e^{Lh_{top}(f_\lambda)}$ \cite{bandt02}, where $h_{top}$ is the topological
entropy of the  map $f_{\lambda}$ \cite{adler}. For unimodal maps the topological entropy
can be easily computed according to the theory of applied symbolic
dynamics \cite{dilao10} and, in some cases, it is even possible to
give a closed analytical form \cite{DeDeus19821}. In
Fig.~\ref{fig:top_log} we show the topological entropy of the logistic
map with respect to the control parameter. According to the scope
depicted by the permutation phases of the cryptosystem defined in
\cite{Wang20121101}, the control parameter should be selected in such
a way that $h_{top}(f_\lambda)$ is greater than $log(2^{52})/(3M)$. If
we consider that the smallest value for $M$ and $N$ is 128, then the
previous restriction is satisfied for $\lambda$ above
$3.57538$. This fact implies a reduction of the key space as defined
in \cite{Wang20121101} and, although it is not a large shortening, it
indeed informs about the needs of using not only the Lyapunov exponent
but also the topological entropy as core of the selection of the keys
of the cryptosystem. 

Finally, another problem when defining the key space of the
cryptosystem arises from the symmetry of the iteration function of the
logistic map. As it is commented in \cite{Li2011949}, the fact that
Eq.~(\ref{eq:logistic}) satisfies $f(x)=f(1-x)$ implies that $x_R$ and
$(1-x_R)$ are equivalent sub-keys for decryption. The same applies to
$x_C$ and $(1-x_C)$.

\subsection{Low sensitivity to the change of plain-image}
\label{sec:low-sens-change}
\begin{figure}[!htbp] \centerline{%
\subfigure[$$]{\includegraphics[scale=0.6]{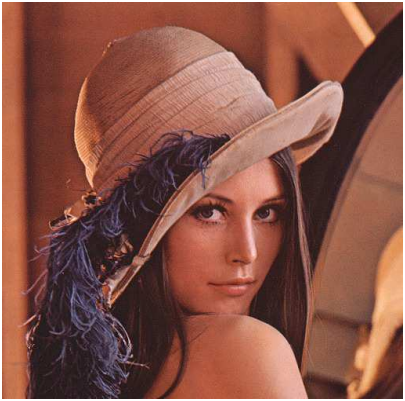}}
\subfigure[$$]{\includegraphics[scale=0.6]{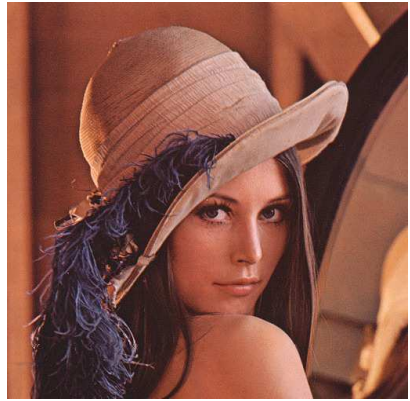}} } \centerline{%
\subfigure[$$]{\includegraphics[scale=0.6]{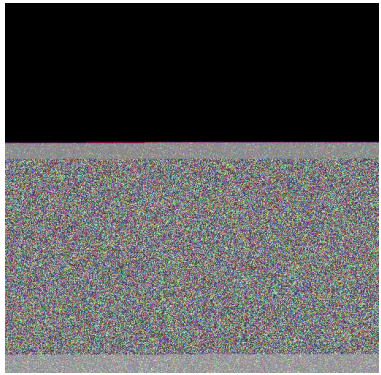}} }
\caption{Example on the low sensitivity to the change of the
plain-image:(a) the first plain-image;(b) first plain-image with the
center pixel of each colour band equals to $255$; (c) XOR between the
cipher-image corresponding to the original plain-image and 
that of the modified one.}
\label{fig:low_sensitivity}
\end{figure}

In the context of cryptography a minor change in the input of a
cryptosystem should imply a major change in the corresponding output
\cite[Rule 9]{Alvarez06a}. In this respect, if we take into account two images
$\mymatrix{I}_0$ and $\mymatrix{I}_1$ with only one different pixel,
then the associated cipher-images should be very different. To
assess this property for the cryptosystem in \cite{Wang20121101}, we have encrypted the images in
Fig.~\ref{fig:low_sensitivity} using as key $\lambda_R=4$,
$x_R=0.1234567898765$, $\lambda_C=3.99$, and
$x_C=0.56789123456789$. The differential cipher-image is equal to zero
for a meaningful set of pixels, which informs about the limitations of
the diffusion property of the cryptosystem given in
\cite{Wang20121101} regarding changes in the plain-image. 

\section{Security analysis: vulnerability against a chosen-plaintext attack}
\label{sec:probl-deriv-from} 
According to \cite[p. 25]{stinson95}, the security assessment of any
cryptosystem must be carried out (at least) with respect to four basic attacks:
\begin{itemize}
  \item Ciphertext-only attack: the cryptanalysis only knows the
    result of encryption.
  \item Known-plaintext attack: several pairs of plaintext and
    ciphertext are accessible for the cryptographer.
  \item Chosen-plaintext attack: the attacker gains access to the
    encryption machine and performs cryptanalysis by selecting
    adequate plaintexts. 
  \item Chosen-ciphertext attack: the decryption machine can be used
    by the cryptanalyst, which chooses ciphertexts in order to extract
    information about the secret parameters of the cryptosystem. 
\end{itemize}
In this section we show that the cryptosystem described in
\cite{Wang20121101} does not exclude the successful application of a chosen-plaintext
attack. 

\subsection{Breaking the confusion stage}
\label{sec:break-conf-stage}
As it has been pointed out in Sec.~\ref{sec:description}, the encryption scheme consists of two classes of
procedures: permutation and substitution of pixels. The main weakness
of the proposal is a consequence of the independence between the
shuffling stages and the last stage, i.e., the one concerning the
substitution of pixels. This fact can be exploited by means of the
following \emph{divide-and-conquer attack}, using as bottom-line
chosen plain-images which are neutral elements with respect to
row/column permutations \cite{arroyo08b}. In this sense, if one
encrypts a plain-image with all pixels equal to the same value, then
the output of the shuffling procedures is the same
plain-image. Moreover, if the plain-image is selected forcing all
rows/columns being equal, then encryption only shuffles columns/rows.

In correspondence to the previous comments, we can mount an attack
based on a chosen plain-image with all pixels equals to zero. Let
$\mymatrix{I}$ be a colour image with all pixels equal to zero, which
implies that $I_S(i)=0$ for $i=1\sim3MN$. Taking into account
Eq.~(\ref{eq:1}), we have
\begin{equation}
  \label{eq:3} I_B(i) = \sum\limits_{j=1}^{i}B(j)\! \!\mod \ 256,
\end{equation} for $i=2\sim 3MN$. From the previous equation we can
find the value of $B(i)$ just by subtracting $I_B(i-1)$ from $I_B(i)$.

If we want to apply the recovered $\mymatrix{B}$ to get any
$\mymatrix{I}_S$ from the corresponding $\mymatrix{I}_B$, then
$\mymatrix{S}$ must be obtained. This commitment can be accomplished
using a second \emph{constant value plain-image}. For instance, we can
use a chosen plain-image with all pixels equal to one. This being the
case, we have $I_S(i)=1$ for $i=1\sim3MN$ and
\begin{equation}
  \label{eq:4} I_B(i) = \sum\limits_{j=1}^{i} (B(j)+ 2j-1) \mod 256,
\end{equation} for $i=1\sim 3MN$. Let us focus on the image given as
the difference between the cipher-images obtained from
$\mymatrix{I}=0$ and $\mymatrix{I}=1$ respectively. Since the
difference between Eq.~(\ref{eq:3}) and Eq.~(\ref{eq:4}) is equal to
$(2j -1)\mod 256$, the components of $\mymatrix{S}$ are determined by
looking for the pixel with value $(2j -1)\mod 256$ in each colour band
of that difference image. If that pixel belongs to the red component,
then $S(j)=0$; if it is one of the green pixels, then $S(j)=1$;
finally, $S(j)=2$ leads to a pixel in the blue band.

\subsection{Permutation-only ciphers}
\label{sec:perm-only-ciph} Once the substitution keystream
$\mymatrix{B}$ and the selection vector $\mymatrix{S}$ have been
obtained, it is possible to reconstruct the input of the shuffling
procedures according to \cite{Li2011949,Li2008212}\footnote{The reader
is referred to these papers for a rigorous study on the security of
permutation-only ciphers. Here the description is limited to the
minimum details required to carry out the implied
cryptanalysis.}. This new goal is going to be achieved by using
$\lceil \log_{256}(3M\times 3MN)\rceil$ chosen plain-images. In this
paper we restrict our analysis to images of the same size as those
used in \cite{Wang20121101}, i.e., images of size $256\times 256$ and,
consequently, four chosen-plain images are required to elude the
permutation-only phase.

In order to validate our cryptanalysis, we have configured an
encryption machine by selecting the key defined by the set $\lambda_R=4$,
$x_R=0.1234567898765$, $\lambda_C=3.99$, and
$x_C=0.56789123456789$. Upon the assumption of having access to the
encryption machine, we encrypt an image equal to zero and an image with
all pixels equal to 1. The cryptanalysis described in
Sec.~\ref{sec:break-conf-stage} is applied, and thus the keystream $\mymatrix{B}$ and the pseudo-random
sequence $\mymatrix{S}$ are recovered. As it is commented in similar
cryptanalysis works \cite{Li2011949,arroyo08b,li09}, the recovering of
those sequences is equivalent to getting the secret key. Nevertheless, the
complete cryptanalysis of the cryptosystem in \cite{Wang20121101}
demands to infer a permutation matrix  representing the composition of
the permutation procedures lead by $\mymatrix{P}_R$ and
$\mymatrix{P}_C$.  This goal can be achieved by using plain-images
with all rows/columns equals. To illustrate the cryptanalysis we are
going to extract the original positions of the  pixels of the first row of 
$\mymatrix{I}^\prime$. First, we encrypt a plain-image with each
colour component determined by 
\begin{equation*}
  \left(\begin{array}{ccccc}
    0& 0& 0& \cdots& 0\\
    1& 1& 1& \cdots& 1\\
    \vdots&\vdots& \vdots&\vdots&\vdots\\
    255& 255& 255& \cdots&255\\
  \end{array}\right)
\label{eq:2}
\end{equation*}
If we consider the vector $\mymatrix{R}_1$ of length 768 given by the concatenation of the first row
of red, green, and blue component of the cipher-image, it is easy to
verify that it contains only three values. The values corresponding to
the selected secret key are 93, 203, and 223, which indicates that the
first row of the cipher image comes from either the row 93, 203, 223
of either of the colour components of the plain image. In order to
establish the colour band of each of the three
candidates for row permutation, we encrypt a plain-image with red
component with all pixels equal to zero, green band being 1, and blue
component being 2. Then,  we look for the occurrences of 0, 1, and 2 in the
first row of each colour component of the cipher-image. The
intersection of this new vector of indexes of occurrence with the
previous one enables to conclude that $\mymatrix{R}_1$ contains the row
93 of the blue
band of the plain-image, the row 203 of the red component of the
plain-image, and the row 223 of the  red component of the
plain-image. After identifying the source of the first row of
$\mymatrix{I}^\prime$, we need to label each pixel of the rows
identified as sources of that row. This aim is fulfilled if we encrypt a colour image with its three
colour components equal to
\begin{equation*}
  \left(\begin{array}{ccccc}
    0& 1& 2& \cdots& 255\\
    0& 1& 2& \cdots& 255\\
    \vdots&\vdots& \vdots&\ddots&\vdots\\
    0& 1& 2& \cdots&255\\
  \end{array}\right)
\end{equation*}
Afterwards,  we look for the occurrences of $i=0\sim 255$ through
the vector $\mymatrix{R}_1$ . The indexes of occurrence are given by
the set $V_i$. Let us begin with $V_0$,  which is $\{120, 356,68\}$
for the selected key. The set $V_0$ implies that either of the referred pixels comes from
the first pixel of either the row 93 of the blue component, the red
row number 203, or the row 223 of the red band  of the
plain-image. To select the proper value among the three candidates for
the three identified pixels, we encrypt an plain-image such that 
the row 93 of the blue component is
\begin{equation*}
\left(0\ 1\ 2\ \cdots \ 253 \ 254 \ 255\right),
\end{equation*}
the red row number 203
\begin{equation*}
  \left(255\ 0\ 1\ 2\ \cdots \ 253 \ 254\right),
\end{equation*}
and the row 223 of the red band is defined as
\begin{equation*}
    \left(254\ 255\ 0\ 1\ 2\ \cdots \ 253\right).
\end{equation*}
Again, we look for $0$ through $\mymatrix{R}_1$  and we obtain the
indexes of occurrence 235, 356, and 556. Only 356 is included in the
previous set $V_0$, and as a result we have that the first pixel of the row
93 of the blue component of $\mymatrix{I}$ goes to the pixel 100
($100\equiv 356\! \mod256$) of
the first row of the green component of $\mymatrix{I}^\prime$. If we
proceed in the same fashion with $V_i$ for $i>0$, then we obtain the
permutations for all the pixels of the row 93 of the blue band of the
plain-image. The same applies to the row 203 (223) of the red band,
but taking into account that the first pixel of the row is now labeled
by 255 (254). 

\begin{figure}[!htbp]
  \centerline{%
    \subfigure[]{\includegraphics[scale=0.6]{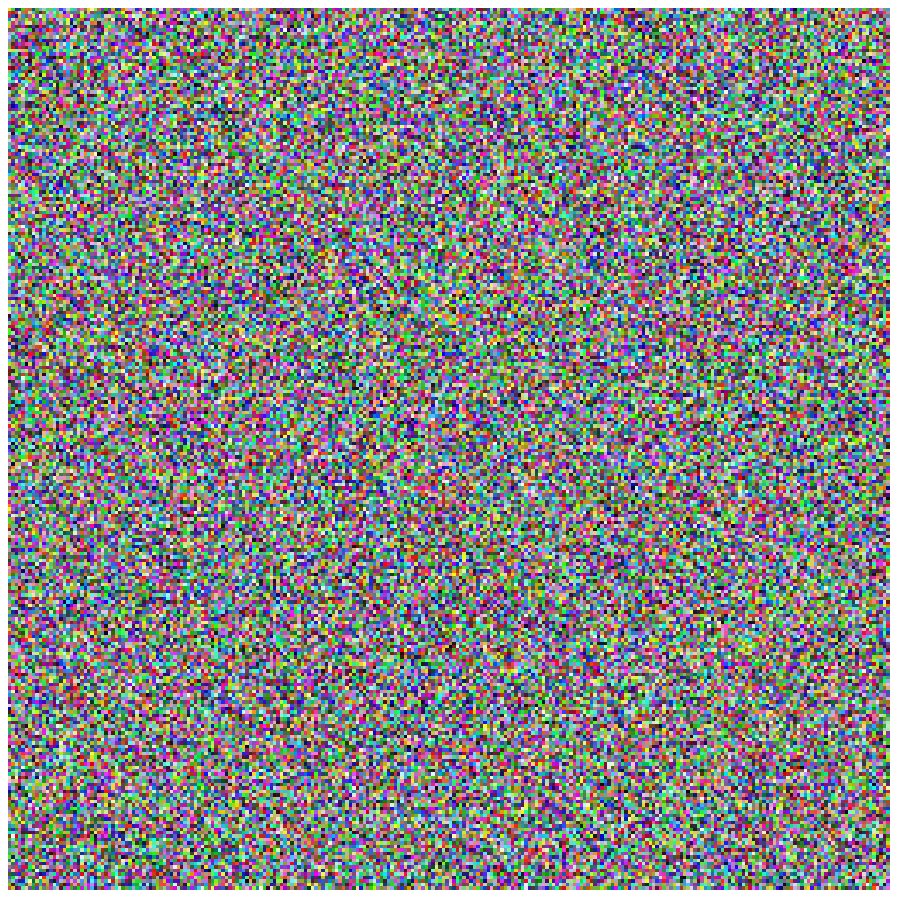}}
    \subfigure[]{\includegraphics[scale=0.6]{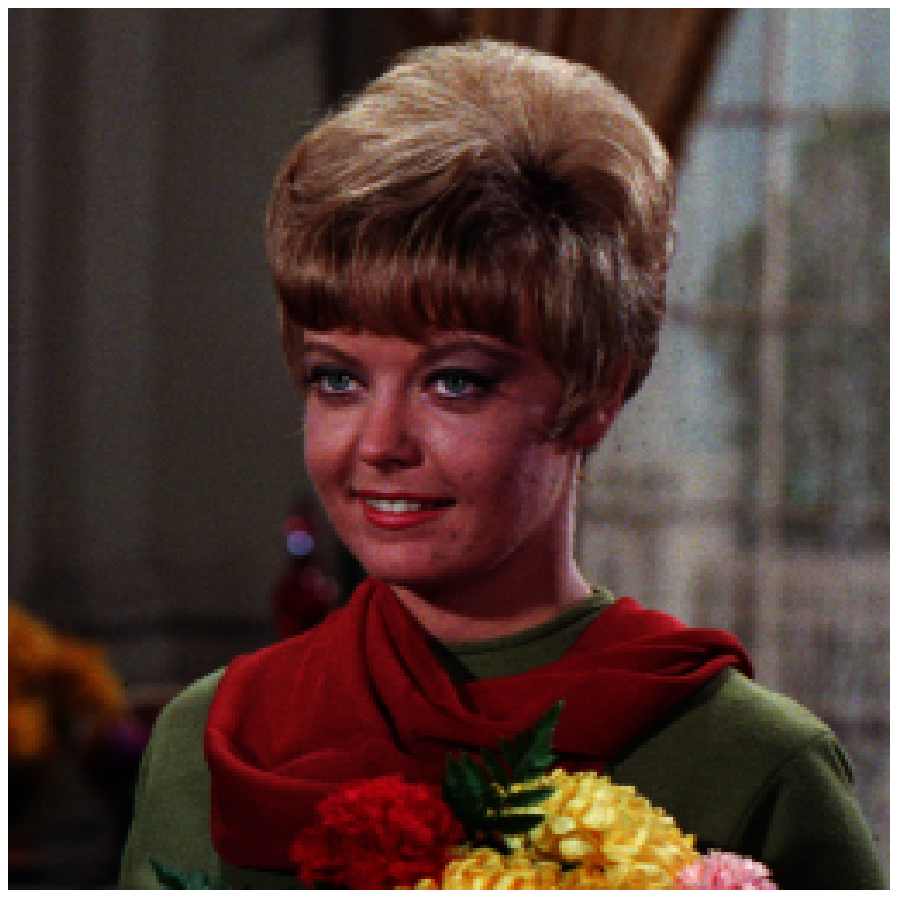}}
  }
  \caption{Application of the chosen-plaintext attack:(a) a
    cipher-image obtained using $\lambda_R=4$,
$x_R=0.1234567898765$, $\lambda_C=3.99$, and
$x_C=0.56789123456789$; (b) the decrypted plain-image using the
keystreams and the permutation matrix inferred via the
chosen-plaintext attack.}
  \label{fig:chosen}
\end{figure}

If one applies the previous methodology for all the rows of the
cipher-image, then the permutation matrix can be inferred. In this
sense, we have applied the cryptanalysis based on the six chosen
plain-images to an encryption machine with secret key $\lambda_R=4$,
$x_R=0.1234567898765$, $\lambda_C=3.99$, and
$x_C=0.56789123456789$.  The cryptanalysis allows to get
$\mymatrix{S}$, $\mymatrix{B}$, and the permutation matrix, which is
equivalent to obtain the secret key. To verify this assertion we have
encrypted an image (the result is in Fig.~\ref{fig:chosen}(a)),
applied the cryptanalysis, and decrypted the cipher-image using the outputs of the cryptanalysis. The decrypted
image is the one in Fig.~\ref{fig:chosen}(b), which coincides with the
original plain-image.

\section{Conclusions}
\label{sec:conclusions} In this paper we have studied in detail a
recent proposal in the area of chaos-based cryptography. We have
underlined some problems related to the dynamical properties of the
system sustaining encryption, and we have also pinpointed some flaws
related to the encryption architecture. The goal of our work was not
simply to show the problems of a given chaotic cryptosystem, but to
highlight the possibility of creating secure proposals to encrypt
information using chaos. In this flavour, our recommendation is on the
line of the set of rules given in
\cite{Alvarez06a,arroyo:thesis,book}.

\section*{Acknowledgments} 

This work was supported by the UAM projects
of Teaching Innovation and the Spanish Government projects
TIN2010-19607 and BFU2009-08473. The work of David Arroyo was
supported by a Juan de la Cierva fellowship from the Ministerio de
Ciencia e Innovaci\'on of Spain.

\bibliographystyle{elsarticle-num}

\end{document}